\documentclass[12pt]{article}
\usepackage{epsfig}
\usepackage{lineno}  
\usepackage{xspace} 
\def\beq{\begin{equation}}
\def\eeq#1{\label{#1}\end{equation}}
\def\eeqn{\end{equation}}

\def\beqa{\begin{eqnarray}}
\def\eeqa#1{\label{#1}\end{eqnarray}}
\def\eeqan{\end{eqnarray}}

\let\bar=\overbar

\def\Dslash{\not{\hbox{\kern-4pt $D$}}}
\def\dslash{\not{\hbox{\kern-2pt $\del$}}}

\def\BR{\mbox{\rm BR}}

\def\msb{{\bar{\ssstyle M \kern -1pt S}}}

\def\Title#1{\begin{center} {\Large {\bf #1} } \end{center}}
\def\Pb      {\ensuremath{b}\xspace}

 \def\bquark    {\ensuremath{\Pb}\xspace}
 \def\bquarkbar {\ensuremath{\overline \bquark}\xspace}
\def\bbbar     {\ensuremath{\bquark\bquarkbar}\xspace}
\def\invfb   {\ensuremath{\mbox{\,fb}^{-1}}\xspace}
\def\sqs   {\ensuremath{\protect\sqrt{s}}\xspace}
\def\cmmd  {\ensuremath{{\rm \,cm}^{-2}}\xspace}
\def\smu  {\ensuremath{{\rm \,sec}^{-1}}\xspace}

\def\Dz      {\ensuremath{D^0}\xspace}
\def\Dbar    {\kern 0.2em\overline{\kern -0.2em D}{}\xspace}
\def\Dzb     {\ensuremath{\Dbar^0}\xspace}

\def\Bd      {\ensuremath{B^0}\xspace}
\def\Bs      {\ensuremath{B^0_s}\xspace}

\def\ra                 {\ensuremath{\rightarrow}\xspace}
\def\Kstarz  {\ensuremath{K^{*0}}\xspace}
\def\KS    {\ensuremath{K^0_{\scriptscriptstyle S}}\xspace} 

\newcommand{\gev}{\ensuremath{\mathrm{\,Ge\kern -0.1em V}}\xspace}
\newcommand{\gevc}{\ensuremath{{\mathrm{\,Ge\kern -0.1em V\!/}c}}\xspace}
\def\mum  {\ensuremath{\,\mu\rm m}\xspace}
\newcommand{\gevdcc}{\ensuremath{{\mathrm{\,Ge\kern -0.1em V^2\!/}c^4}}\xspace}

\begin{document}

\begin{flushright}  
LHCb-PROC-2013-005
\bigskip\bigskip
\end{flushright}

\Title{Future prospects at LHCb}

\bigskip\bigskip


\begin{raggedright}  

{\it Marie-H\'el\`ene Schune\footnote{On behalf of the LHCb collaboration.}\index{Schune, M.-H.}\\
LAL, Universit\'e Paris-Sud \\
CNRS/IN2P3\\
Orsay, FRANCE}
\bigskip\bigskip
\end{raggedright}

\begin{abstract}
The LHCb experiment is running at the Large Hadron Collider to study CP violation and rare decays in the beauty and charm sectors. The motivation and the strategy of the upgrade envisaged for the long shutdown LS2 (2018) is presented. The current results for some exemplary physics analyses are given and the expected performances foreseen for 2018 and for the LHCb Upgrade project with an integrated luminosity of 50~\invfb are summarized.
\end{abstract}

\section{Introduction}
The LHCb experiment, located on the Large Hadron Collider (LHC), has been successfully taking data during the last three years in running conditions very encouraging for the future. The general context of the LHCb Upgrade project is presented, followed by a short overview of the main changes foreseen. Finally, using some exemplary physics analyses, the current results are shown and the expected precisions are given. 

\section{General context}
\subsection{Introduction}
\begin{figure}[hbtp]
\begin{center}
{\includegraphics[height=10cm]{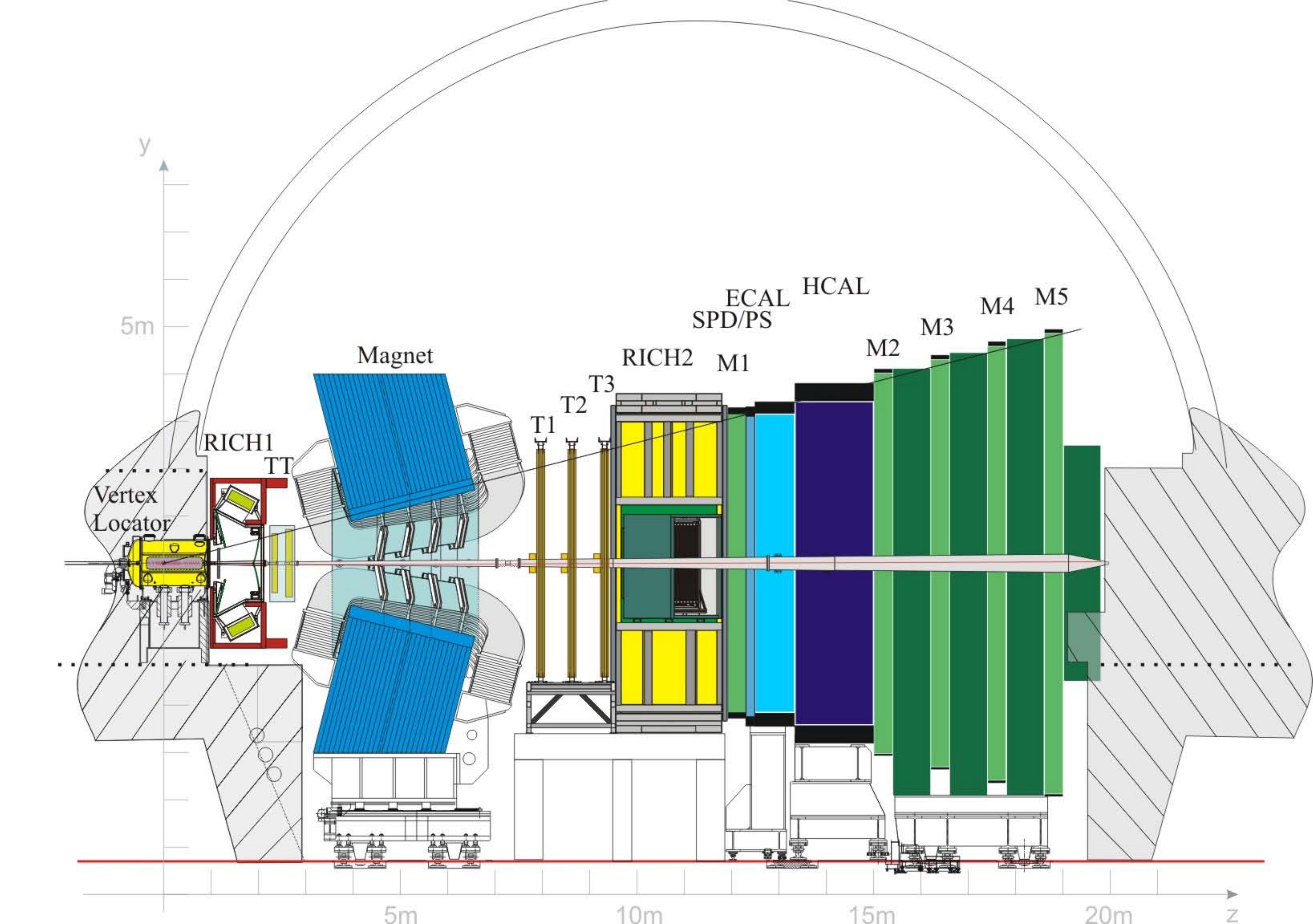}}
\caption{Vertical view of the LHCb detector.}
\label{fig:LHCb}
\end{center}
\end{figure}

Studies of CP violation and more generally flavor changing neutral currents were essential in establishing the Standard Model (SM). 
Today, they appear as a powerful tool to reveal processes beyond the SM and to understand their nature. In this context, LHCb plays a key role.
The LHCb detector\cite{bib:LHCb}, shown in Fig.\ref{fig:LHCb}, is a single-arm forward spectrometer covering the \mbox{pseudorapidity} range $2<\eta <5$, designed
for the study of particles containing $b$ or $c$ quarks. The detector geometry is driven by the kinematics of the
\bbbar pair production at the LHC energy where both \bquark and \bquarkbar quarks mainly fly in the forward or backward direction.
The
detector includes a high precision tracking system consisting of a
silicon-strip vertex detector surrounding the $pp$ interaction region,
a large-area silicon-strip detector located upstream of a dipole
magnet with a bending power of about $4{\rm\,Tm}$, and three stations
of silicon-strip detectors and straw drift tubes placed
downstream. The combined tracking system has a momentum resolution
$\Delta p/p$ that varies from 0.4\% at 5\gevc to 0.6\% at 100\gevc,
and an impact parameter resolution of 20\mum for tracks with high
transverse momentum. Charged hadrons are identified using two
ring-imaging Cherenkov detectors. Photon, electron and hadron
candidates are identified by a calorimeter system consisting of
scintillating-pad and preshower detectors, an electromagnetic
calorimeter and a hadronic calorimeter. Muons are identified by a
system composed of alternating layers of iron and multiwire
proportional chambers. The trigger consists of a hardware stage, the Level-0 (L0) trigger based
on information from the calorimeter and muon systems, followed by a
software stage the High Level Trigger (HLT) which applies a full event reconstruction.
\subsection{Current data taking and longer term plans}
During 2011, both the LHC machine and the LHCb detector performed superbly. This has allowed LHCb to accumulate 1.0 \invfb of \sqs = 7 TeV $pp$ collisions that is available for physics analysis. The detector was running at an instantaneous luminosity of about $3.5 \times 10^{32}$\cmmd \smu , and with a number of $pp$ interactions per crossing of $\sim 1.4$. The values of these running parameters are above the design values by a factor 1.8 and 3.5 respectively. The large integrated luminosity was collected thanks to the luminosity leveling schema set in place by the LHC team. Instead of being defocused, the beams are displaced from head-on collision in order to achieve the required luminosity.  This offset is reduced with time in order to follow the loss in beam intensity throughout the fill.   This procedure allows a constant luminosity to be maintained during the whole period of the fill, in contrast to ATLAS and CMS which experience an exponential decrease.
\par
During 2012 data taking, the center of mass energy has been increased by $\sim 15 \%$ (  \sqs = 8 TeV) which corresponds to an increase of about  15 \% in the number of \bbbar  events. 
The  instantaneous luminosity is also slightly larger : $4 \times 10^{32}$\cmmd \smu .  About 1.4~\invfb of $pp$ collisions have been already recorded and three months of data taking are still planned.\footnote{At the end of the 2012 run 2 \invfb of $pp$ collisions at \sqs = 8 TeV have been recorded.} In addition, two changes have been made for 2012 data taking : the HLT output rate has been increased from 3 kHz in 2011 to 4.5 kHz which mainly allows to record more charm events and to cope with the increase in cross-sections. A  deferred trigger schema has also been put in place. In this schema about 20 \% of the events are written to disk before the HLT decision which is then made during the inter-fill gap periods. It results in an increase of about 20\% in the data sample. These two novelties ensure a larger data sample to work with during the two years shut-down period which will take place in 2013-2014.   
\par
After this shutdown period, a three years data taking period is foreseen at a center of mass energy of  \sqs = 13-14 TeV which corresponds approximately to a doubling of the \bbbar cross section compared to 2011. It is expected that the recorded integrated luminosity during this period should be of the order of 5~\invfb . 
After these three years, according to the LHC planning, a second shutdown (LS2, starting in 2018) will take place and the upgraded LHCb detector will be installed. 
Very important changes in the detector will be made (see Sec.~\ref{sec:Upgrade}) and it is foreseen to be able to record 5~\invfb per year for a total of 50\invfb at a center of mass energy of 14 TeV. 
\par 

\subsection{Why do we need to upgrade the LHCb experiment ?}
Using the 2011 dataset, the LHCb experiment has already been able to perform extremely precise measurements. In most cases the precision obtained is the world's best. Unfortunately, all the measurements are consistent with the SM and  New Physics (NP) has not shown up. However, even if the measurements are much more precise than before, there is still plenty of room for NP. As an example, let's take the global fits as performed by the CKMFitter \cite{bib:CKMFitter} or UTFit \cite{bib:UTFit} collaborations. In these fits they allow for the presence of generic NP signals in the \Bd or \Bs mixing. Whatever the statistical treatment, there is a good agreement with the SM but, due to the size of the uncertainties, the presence of NP at the level of 20 to $30 \%$ is still possible. It is thus highly desirable to largely increase the available data samples and this can only be achieved through redesign of the LHCb experiment both from the detector point of view, to be able to cope with a larger instantaneous luminosity, but also from a trigger point of view as will be explained in the next part.    

\section{The LHCb upgrade \label{sec:Upgrade}}
The aim of the upgraded LHCb experiment  is to reach experimental sensitivities comparable to or better than the theoretical ones and thus to record an integrated luminosity of about 
50~\invfb. It implies
to run at ${\cal L} = 1 \times 10^{33}$ \cmmd \smu with a fully flexible software trigger running at a readout rate of 
40~MHz~\cite{bib:Upgrade-LoI,bib:Upgrade-FTDR}. This increases the annual signal yields, as compared to those obtained by LHCb in 2011, by a factor around ten for muonic B decays and twenty or more for heavy-flavour decays to hadronic final states. For reason of flexibility, and to allow for possible evolutions of the trigger, it was decided to design those dectectors that need replacement such that they can sustain a luminosity of ${\cal L} = 2 \times 10^{33} \cmmd \smu$.
The challenge of this update is related to the number of $pp$ interactions per crossing. The average number of interactions per crossing is about 2.3 at ${\cal L} = 1 \times 10^{33} \cmmd \smu$ and reaches more than 4 interactions per crossing at ${\cal L} = 2 \times 10^{33} \cmmd \smu$. In the latter case, all crossings have at least one interaction. These extreme conditions put heavy requirements on the tracking as well as on the trigger algorithms. 

\subsection{The current LHCb trigger}
The current LHCb trigger~\cite{bib:LHCbTrigger} contains two stages, the L0  and the HLT. The L0 reduces the rate from 40~MHz down to 1~MHz. It is based on custom electronics receiving dedicated information from the calorimeters and from the muon detector. It looks for lepton and hadron candidates with a high transverse momentum.
The HLT trigger reduces the rate down to 3 to 4.5~kHz. The HLT is a software trigger running on a dedicated CPU farm and receiving the full detector information at 1~MHz. By running tracking and vertexing algorithms it selects leptons and hadrons with a high transverse momentum as well as a high impact parameters. More elaborate algorithms, close to the off-line selections, are then applied to select inclusive or exclusive heavy-quark decays.
The L0 saturates for hadronic channels when the luminosity increases~\cite{bib:Upgrade-LoI}. At high luminosity we cannot rely only on the transverse momentum cut for efficient triggering. On the contrary, a software based trigger allows use of many discriminants including track impact parameters and combinations of different criteria.

\subsection{The upgraded LHCb trigger}
The main point in the upgraded LHCb trigger is to remove the 1~MHz bottleneck at the output of the L0 trigger which is determined by the front-end electronics rate. These front-end electronics will be upgraded to allow reading events at the LHC clock rate. 
 In principle, in the upgrade schema, the data acquisition (DAQ) and event building could be performed in a large CPU farm at the full rate of 40~MHz. However the upgrade is designed to be able to cope with a staged DAQ system which cannot yet handle the full rate, occupancy fluctuations which prevent the full readout, and insufficient CPU power in the CPU farm. Hence the upgrade will also contain a Low Level Trigger (LLT), which,  similarly to the current L0 trigger,  should not just pre-scale to a rate acceptable, but enrich the selected sample with interesting events. The LLT corresponds closely to the current L0, but with a tunable output rate higher than the current 1~MHz limit. In the case of hadronic channels, such as for example the $\Bs \ra \phi \phi$ decay mode, with a LLT output rate of 10~MHz, the trigger efficiency should be about 4 times larger than with the current 1~MHz rate~\cite{bib:Upgrade-LoI}. 

\subsection{Sub-detectors upgrades}
In addition to the change of most of the front-end electronics, several parts of the current LHCb detector will have to be replaced in order to cope with the increased occupancy  \cite{bib:Upgrade-LoI}. The open geometry of the LHCb detector will allow portions of the upgraded detector to be installed in any reasonably long shutdown.
\par
The VELO is a silicon strip detector with $r$ and $\Phi$ geometry. It will be replaced by either a pixel or a strip detector. The pixel version provides a very low occupancy for each channel, reducing the combinatorial for the tracking algorithm. A decrease of the inner radius is also studied in order to improve the impact parameter resolution which is one of the key ingredients for the HLT. 
The TT stations are a silicon-strip based detector. They will be replaced using the same technology with an enlarged acceptance and an improved granularity.
The T stations are composed of an OT with straw tube detectors and an IT with silicon-strip detectors to cover the high occupancy area near the beam pipe. To account for the higher occupancy due to the increase in luminosity, two options are studied, a large area silicon-strip IT completed by OT straw tubes, or a Central Tracker made from scintillating fibers read by silicon photomultipliers.
\par
The current RICH photo detectors are HPDs in which the readout chip, operating at 1 MHz, is encapsulated within the vacuum tube.  The candidate replacement technology is a multi-anode photomultiplier, read out at 40 MHz. In addition, due to the increased occupancy in the detector, the aerogel part of the RICH-1 will be removed. Studies are on-going to use a novel detector based on time-of-flight to identify low momentum particles below $\sim 10 \gevc$~\cite{bib:LHCbTORCH}. 
\par
The calorimeter upgrade consists in replacing the current front-end electronics by a new system able to send data to the DAQ at 40~MHz. 
Moreover, the new electronics must have a gain five times higher than the present system, in order to compensate for a gain reduction that will be imposed on the photomultipliers so that the mean anode current remains at an acceptable level during high-luminosity running.
The muon systems will almost remain unchanged, but the M1 station will be removed. 

\section{Upgrade prospects for some selected physics channel \label{sec:Physics}}
The current precisions for few selected channels are given and further extrapolated to the expected precision in 2018 (before the LHCb Upgrade) and after the LHCb Upgrade phase when an integrated luminosity of 50~\invfb will have been recorded. Obviously, because of time constraints, many interesting analyses are not discussed. In particular the whole field of CP violation in the \Bs has been omitted. The $\phi_s$ and $\Delta \Gamma_s$ observables have been determined simultaneously from $\Bs \ra J/\psi \phi$ decays using using time-dependent flavour tagged angular analyses~\cite{bib:LHCb_phis}. It is expected that the determination of $\phi_s$ will remain statistics limited, even with the data samples available after the upgrade of the LHCb detector~\cite{bib:ImplicationDoc}.  
  
\subsection{The $\Bs \ra \mu^+ \mu^-$  and $\Bd \ra \mu^+ \mu^-$ decays}  
\begin{figure}[htpb] 
\begin{center}
{\includegraphics[height=4.5cm]{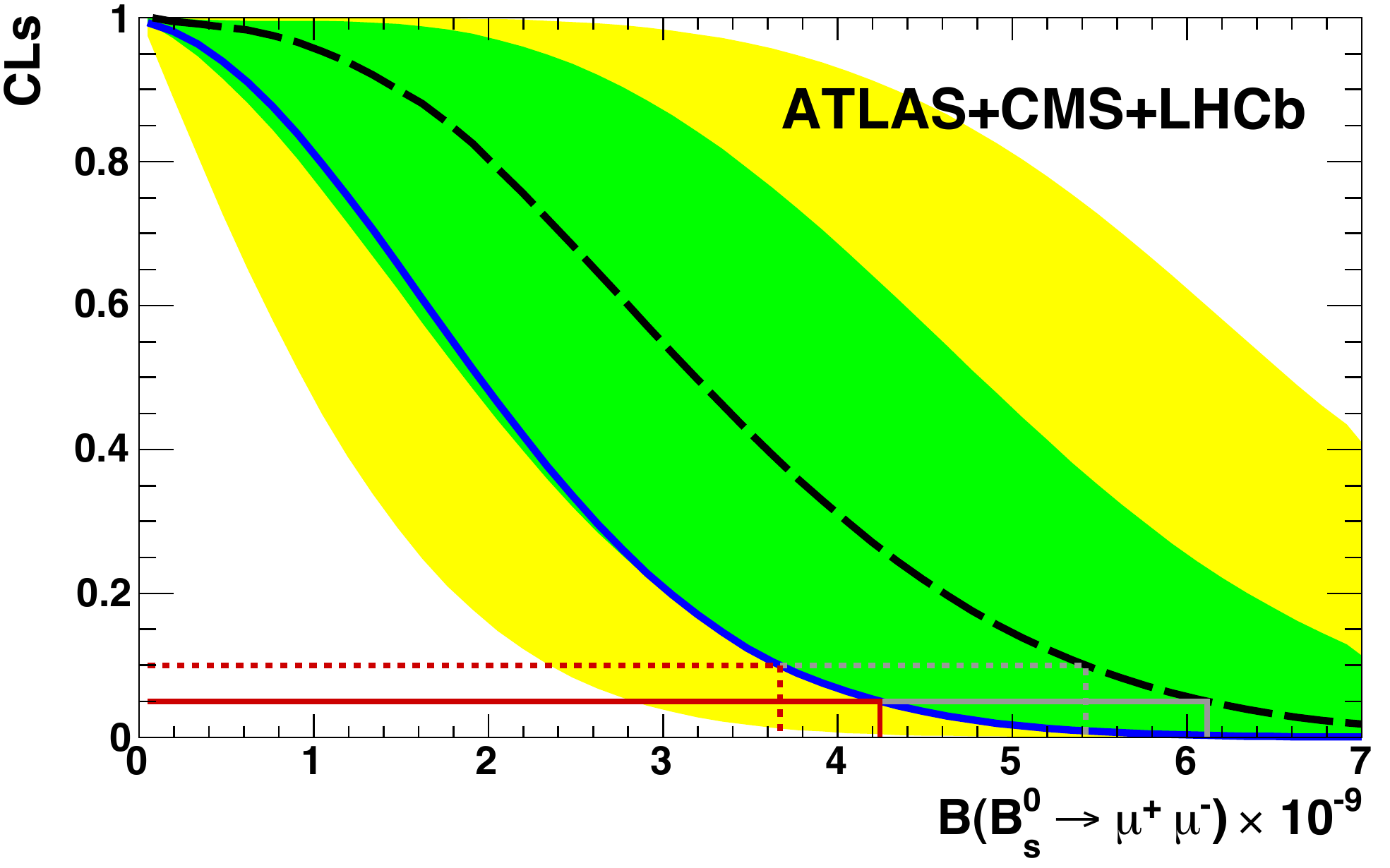}}
{\includegraphics[height=4.5cm]{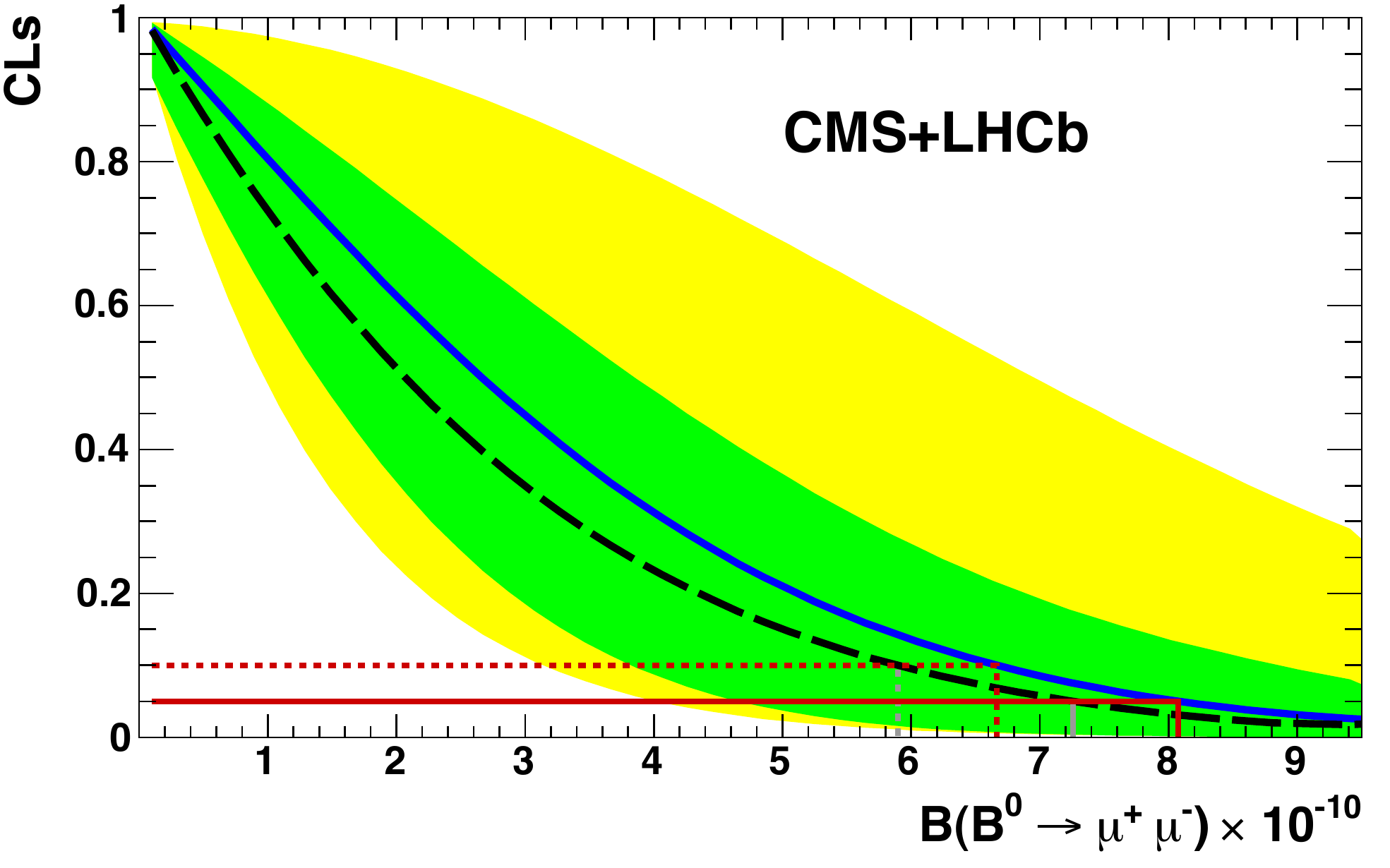}}
\caption{CLs as a function of the assumed branching fraction for (left) $\Bs \ra \mu^+ \mu^-$  and (right) $\Bd \ra \mu^+ \mu^-$ decays. The long dashed black curves are the medians of the expected CLs distributions for $\Bs \ra \mu^+ \mu^-$, if background and SM signal were observed, and for $\Bd \ra \mu^+ \mu^-$,  if background only was observed. The yellow areas cover, for each B, 34\% of the expected CLs distribution on each side of its median. The solid blue curves are the observed CLs. The upper limits at 90\% (95\%) C.L. are indicated by the dotted (solid) horizontal lines in red (dark gray) for the observation and in gray for the expectation.}
\label{fig:mumu}
\end{center}
\end{figure}

\begin{figure}[hbtp]
\begin{center}
{\includegraphics[height=9cm]{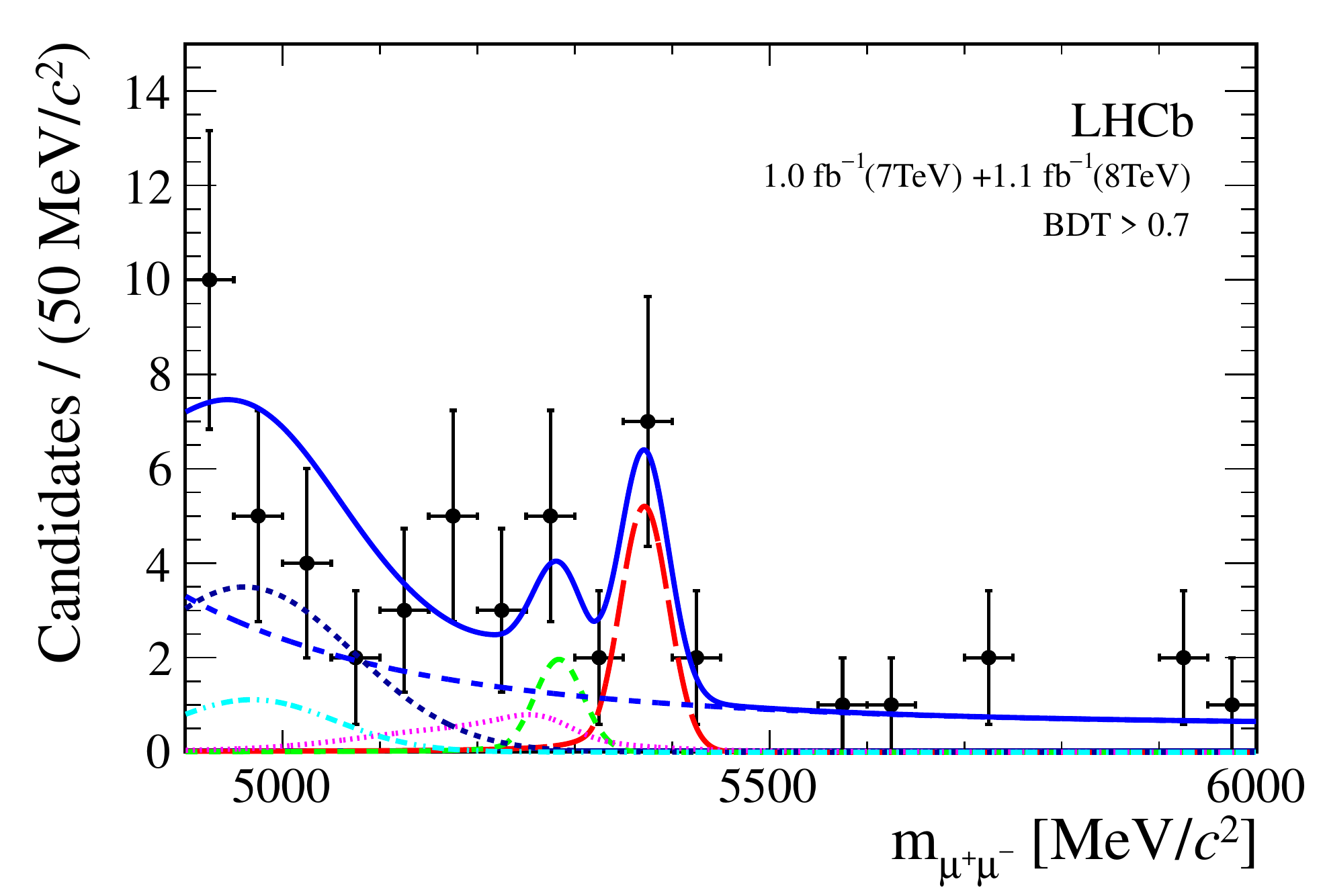}}
\caption{Invariant mass distribution of the selected $\Bs \ra \mu^+ \mu^-$  candidates (black dots) in a signal enriched region. The result of the fit is overlaid (blue solid line) and the different component detailed : $\Bs \ra \mu^+ \mu^-$  (red long dashed),  $\Bd \rightarrow \mu^+ \mu^-$ (green medium dashed), $B^0_{(s)} \rightarrow h^+ h^{'-}$ (pink dotted), $\Bd \rightarrow \pi^- \mu^+ \nu_{\mu}$ (black short dashed) and $B^{0(+)} \rightarrow \pi^{0(+)} \mu^+ \mu^-$  (light blue dot dashed), and the combinatorial background (blue medium dashed).}
\label{fig:Rare_BsToMuMu_LHCb}
\end{center}
\end{figure}

Precise measurements of very rare decays can give some insights of 
the presence of NP particles present in the loop or box diagrams. A very good example of such type of decays is
 $\Bs \ra \mu^+ \mu^-$ which, in the SM, occurs due to very suppressed loop diagrams.
Its branching fraction is expected to be extremely small due to the values of the CKM matrix elements, the GIM mechanism and the helicity suppression. However NP models can enhance it very significantly. 
Similarly one can also search for  $\Bd \ra \mu^+ \mu^-$ which is even more suppressed due to the relevant CKM matrix elements. 
The summary of the reach of the LHC experiments \cite{bib:Bs2MuMu} is shown in Figure~\ref{fig:mumu}.
The obtained 95 \% CL limits obtained using the CL$_s$ method~\cite{bib:CLs} are $4.2 \mbox{ } 10^{-9}$ for $\Bs \ra \mu^+ \mu^-$  
and $ 8.1  \mbox{ } 10^{-10}$ for $\Bd \ra \mu^+ \mu^-$. This corresponds to about 1.2 times the SM value for the \Bs meson and 
still 8 times the SM value for the \Bd meson. In case of the \Bs meson the number of observed candidates is consistent with the expected number of events taking into account the background and assuming the SM value for the branching ratio.  
This already put severe constraints on New Physics models, such as the constrained MSSM for large values of the $\tan \beta$ parameter as can be seen for example in \cite{bib:Nazilla}. Very recently the LHCb experiment has announced the first evidence for the  $\Bs \ra \mu^+ \mu^-$  
decay~\cite{bib:BsToMuMu_LHCb_evidence} leading to :
\begin{equation}
\BR(\Bs \ra \mu^+ \mu^- ) = (3.2 ^{+1.5}_{-1.2}) \times 10^{-9}
\end{equation}
in agreement with the SM prediction.  The invariant mass distribution of the 
$\Bs \ra \mu^+ \mu^-$  candidates in a signal enriched region is shown in Fig. \ref{fig:Rare_BsToMuMu_LHCb}. 

The expected precisions on $\BR(\Bs \ra \mu^+ \mu^-)$ and on $\BR(\Bd \ra \mu^+ \mu^-)/\BR(\Bs \ra \mu^+ \mu^-)$, which is an interesting probe of Minimal Flavor Violation models, are summarized in Table~\ref{tab:BsToMuMu} and compared to the theory uncertainties. 
\begin{table}[htbp]
\begin{center}
\caption{Comparison of the expected precisions as obtained or foreseen by the LHCb  experiment  on 
$\BR(\Bs \ra \mu^+ \mu^-)$ and on $\BR(\Bd \ra \mu^+ \mu^-)/\BR(\Bs \ra \mu^+ \mu^-)$.}
\vskip .5cm
\begin{tabular}{l|c|c|c|c}
\hline
										&LHCb 2012 	& LHCb 2018 	& LHCb-Upgrade 			& Theoretical \\
										&			& 		 	& 		 (50 \invfb)		&  uncertainty \\
\hline
$\BR(\Bs \ra \mu^+ \mu^-$) 					& $1.5 \ 10^{-9}$& $0.5 \ 10^{-9}$&$0.15 \ 10^{-9}$			& $0.3 \ 10^{-9}$ \\
$\frac{\BR(\Bd \ra \mu^+ \mu^-)}{\BR(\Bs \ra \mu^+ \mu^-)}$ & \--  & 100 \%								&$\sim 35 \%$ &$\sim 5 \%$ \\
\hline
\end{tabular}
\label{tab:BsToMuMu}
\end{center}
\end{table}

Measuring the branching fractions are not the only interesting physics results one can get from the analysis of the $\Bs \ra \mu^+ \mu^-$  and $\Bd \ra \mu^+ \mu^-$ decays. For example the measurement of the $\Bs \ra \mu^+ \mu^-$ effective lifetime is envisaged as it is a clean test of NP. With the upgraded LHCb detector  this appears to be feasible~\cite{bib:TauEff_BsToMuMu}.

\subsection{The $\Bd \ra \Kstarz \mu^+ \mu^-$  decay} 
The decay $\Bd \ra \Kstarz \mu^+ \mu^-$ is a flavour changing neutral current process which proceeds via electroweak loop and box diagrams in the SM. 
Physics beyond the SM, such as SUSY, can contribute with a comparable amplitude via gluino or chargino loop diagrams and modify observables in the three-body final state. A number of angular observables in $\Bd \ra \Kstarz \mu^+ \mu^-$  decays can be predicted as functions of the dimuon invariant mass squared, $q^2$. These include the forward-backward asymmetry of the muons, $A_{FB}$ or the fraction of longitudinal polarization. The CP asymmetry is also an interesting probe of NP~\cite{bib:KstarMuMu_Theory}. The $\Bd \ra \Kstarz \mu^+ \mu^-$   decay is normally treated using the framework of the operator product expansion, where the Wilson coefficients 
$C_{7,9,10} $ dominate~\cite{bib:KstarMuMu_TheoryWilson}. These have right-handed partners $C^{'}_{7,9,10} $, that are highly suppressed in the SM and in minimal flavour-violating models~\cite{bib:KstarMuMu_ThoeryKM}. In the presence of NP, the values of these coefficients may change due to new heavy degrees of freedom in the loops. Measuring the Wilson coefficients then allows for entire classes of NP to be observed or excluded.
\par
\begin{figure}[htpb]
\begin{center}
{\includegraphics[height=10.0cm]{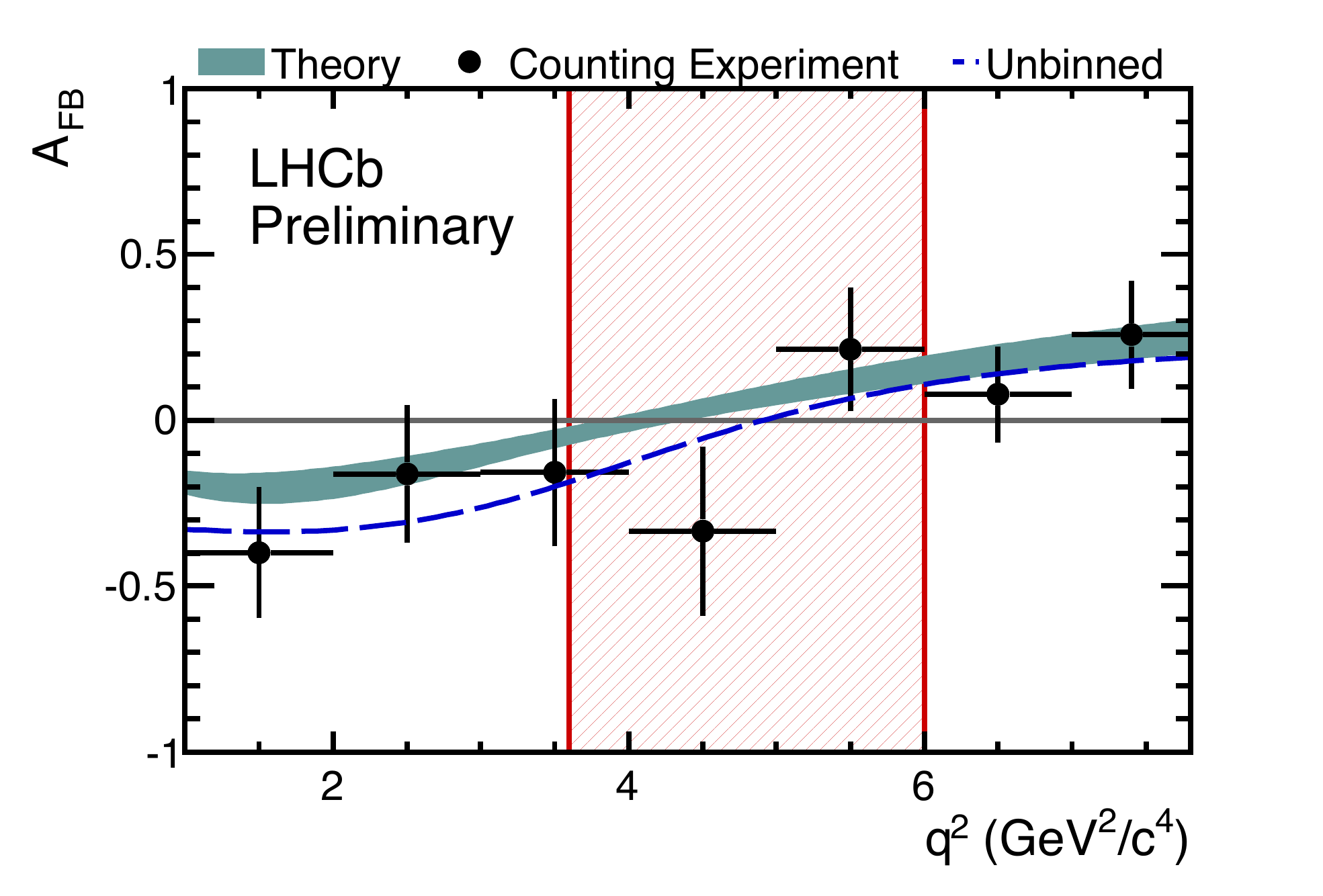}}
\caption{$A_{FB}$ as a function of $q^2$, that comes from an unbinned counting experiment (blue dashed line) overlaid with the theory prediction.
The data-points are the result of counting forward- and backward-going events in 1 \gevdcc bins of $q^2$. The uncertainty on the data points is statistical only. The red-hatched region is the 68\% confidence interval on the zero-crossing point observed in the data.}
\label{fig:AFB_qz}
\end{center}
\end{figure}

The measurement of several angular variables as a function of $q^2$ has been performed recently by the LHCb experiment using 1.0~\invfb\ 
of data~\cite{bib:KstarMuMu_LHCb}. As an example the $q^2$ value at which $A_{FB}$ is 0 can be seen on Fig.~\ref{fig:AFB_qz} with the theory prediction from~\cite{bib:KstarMuMu_ThPlots} underlayed. This is of particular interest since it is both sensitive to NP and is cleanly predicted in the SM. The measured value is
\begin{equation}
 q^2_0= (4.9 ^{+1.1}_{-1.3}) \ \gevdcc 
\end{equation} 
in good agreement with the SM.  In 2018 the expected precision from LHCb is of the order of 6\% and should reach 2\% using the LHCb Upgrade data sample (50~\invfb).
But the forward-backward asymmetry is not the only interesting observable and important theoretical work has been going on to design variables which are at the same time free of hadronic form factor uncertainties and experimentally accessible. Some examples are given 
in~\cite{bib:KstarMuMu_Obs1,bib:KstarMuMu_Obs2}. One such variable is $ A_T^2 (q^2)$ which, at low $q^2$, is sensitive to
 right-handed coupling due to the
$C^{'}_{7} $ Wilson coefficient.  The expected experimental uncertainties are of the order of 0.1 in 2018 and should reach 0.03 using the LHCb Upgrade data sample (50~\invfb). However, in order to reach this precision, subtle effects will have to be taken into account such as the S-wave contribution below the \Kstarz\ peak or threshold effects at  low $q^2$ due to the non-zero muon mass. 
\par 
In parallel with the increasing experimental precision, important phenomenological work is recently on-going to constrain directly the Wilson coefficients (see for example~\cite{bib:KstarMuMu_W1,bib:KstarMuMu_W2}) using many $b\ra s \gamma$ related decays. All the results are currently in good agreement with the SM. However large bins in $q^2$ are still used and, with more statistics and thus finer binning, as will be provided by the LHCb Upgrade experiment, the sensitivity will be greatly improved. 

\subsection{The angle $\gamma$}
In $B$ hadron decays, tree level processes are generally expected to be dominated by SM contributions, while NP is mainly expected to affect loop diagrams. Any difference between the CKM unitarity triangle as measured in tree level processes compared to loop level processes would therefore be an indicator of NP.
\par
The last example briefly presented here is the measurement of the $\gamma$ angle of the Unitarity Triangle  exploiting the interference between $b \ra c$ and $b \ra u $ transitions and using tree-mediated diagrams. All the decay modes which can be used to constrain $\gamma$ are fully hadronic decay modes. As such they will greatly benefit from the upgrade of the LHCb trigger.  The most common measurements involve the decay $B^+ \ra D K^+$. The intermediate $D$ meson can either be a \Dz or a \Dzb and the final state of the $D$ meson is accessible from either flavour state. 
Three methods named following the original suggestion of their use have been pursued by the LHCb experiment on 2011 data (1.0~\invfb) :
\begin{itemize}
\item GLW~\cite{bib:GLW} where the $D$ meson decays to a CP eigenstate : $D \ra KK$ or $D \ra \pi \pi$ 
\item ADS~\cite{bib:ADS} where the $D$ meson decays either to a Cabibbo-favoured or a doubly-Cabbibo-suppressed  final state : 
$D \ra K\pi$ or $D \ra K \pi \pi \pi$ 
\item GGSZ~\cite{bib:GGSZ} where the $D$ meson decays to self-conjugate multibody modes : 
$D \ra \KS \pi \pi $ or $D \ra  \KS K K$ 
\end{itemize}
The power of all these measurements is realized in the combination of all of them giving~\cite{bib:Gamma_LHCb} : 
\begin{equation}
\gamma = (71.1 ^{+16.6}_{-15.7}) ^{\circ}
\end{equation}
which is already competitive with the B-factories average.
The expected precision is foreseen to be of the order of  $4 ^{\circ}$ in 2018 and should reach  $0.9 ^{\circ}$ using the LHCb Upgrade data sample~\cite{bib:ImplicationDoc}.  
The details of the expected precision for the various decay modes with 50~\invfb  as well as the current status are given in Table~\ref{tab:Gamma}. The foreseen precision will be enough to perform a meaningful comparison with the $\gamma$ measurements obtained from decays involving loop diagrams. 
 In addition to those modes, several others are being currently studied 
($\Lambda_b$ decays, $\Bs \ra D \phi$ decays \dots). 
Some of them are challenging in the LHC environment due to the large number of low transverse momentum tracks and the 1~MHz limitation of the current L0 trigger imposing a high (of the order of 3.5 \gev )  transverse energy cut. They will benefit a lot from the upgraded trigger. 
\begin{table}[htbp]
\begin{center}
\caption{Summary of  $\gamma$ measurements for various charmed $B$ decay modes. The current status and the estimated precision with 50~\invfb  are given.}
\vskip .5cm
\begin{tabular}{l|c|c}
\hline
Decay mode		& Cuurent status & $\gamma$ sensitivity \\
	&  (1.0 \invfb)	& (50 \invfb)\\
\hline						
$B \ra DK$, $D \ra hh^{'}$, $D \ra K \pi \pi \pi$ & first CP measurements done 	&		$1.3 ^{\circ}$		\\
$B \ra DK$, $D \ra \KS  hh$	 & first CP measurements done &	$1.9 ^{\circ}$			\\
$B \ra DK$, $D \ra 4 \pi$	& under study &	$1.7 ^{\circ}$			\\
$B \ra DK \pi$, $D \ra hh^{'}$, $D \ra \KS  hh$	 & first CP measurements done &	$1.5 ^{\circ}$			\\
									 & ($B \ra D \Kstarz$, $D \ra hh^{'}$)&			\\
$B \ra DK \pi \pi$, $D \ra hh^{'}$	& first CP measurements done 	&	$\sim 3 ^{\circ}$			\\
Time dependent $\Bs  \ra D_s K$ 	 & first CP measurements done &	$\sim 3 ^{\circ}$			\\
\hline
Combined 		& 	&$\sim 0.9 ^{\circ}$			\\
\hline
\end{tabular}
\label{tab:Gamma}
\end{center}
\end{table}

\section{Conclusions}
The LHCb experiment is currently taking data successfully at the LHC. After the experiment has run for about five years at about twice its design luminosity it is proposed to upgrade the experiment to run at higher luminosity. The upgrade is being organized to be ready for data-taking after the second long shutdown of the LHC, which is scheduled for 2018. One main focus of the upgrade is to increase the read-out of the experiment to 40~MHz, reducing thus the 1~MHz bottle-neck of the first level trigger which is causing an important loss in efficiency most notable in the hadronic channels. 
The LHCb upgrade, aiming at recording an integrated luminosity of 50~\invfb,  is necessary to take the next step in sensitivity that will be required in flavour physics after the first period of exploration and measurement that the experiment will perform over the coming five years. The performance of the existing detector, and the purity of the samples already accumulated, gives confidence that measurements of very high sensitivity will be possible with these samples.

\end{document}